\documentclass[11pt,a4paper]{article}

\usepackage{jheppub}
\usepackage{amsmath,amssymb,bm,booktabs,array,mathtools}
\usepackage{graphicx}
\usepackage{placeins}
\usepackage{needspace}
\usepackage{xcolor}
\usepackage{microtype}
\usepackage{siunitx}
\def\be{\begin{equation}}
\def\ee{\end{equation}}
\def\bea{\begin{array}}
\def\eea{\end{array}}
\def\beqa{\begin{eqnarray}}
\def\eeqa{\end{eqnarray}}
\def\beqas{\begin{eqnarray*}}
\def\eeqas{\end{eqnarray*}}
\def\bp{\begin{picture}}
\def\ep{\end{picture}}
\def\bc{\begin{center}}
\def\ec{\end{center}}
\def\bfig{\begin{figure}}
\def\efig{\end{figure}}

\def\bit{\begin{itemize}}
\def\eit{\end{itemize}}

\def\f{\frac}

\def\[{\left[}
\def\]{\right]}
\def\({\left(}
\def\){\right)}

\def\..{\left.}
\def\.{\right.}

\def\tm{\times}

\makeatletter
\gdef\@fpheader{Prepared for submission to JHEP}
\makeatother

\newcommand{\GeV}{\,\mathrm{GeV}}
\newcommand{\TeV}{\,\mathrm{TeV}}
\newcommand{\keV}{\,\mathrm{keV}}
\newcommand{\MeV}{\,\mathrm{MeV}}
\newcommand{\dmzero}{\delta m_0}
\newcommand{\MPl}{M_{\rm Pl}}
\newcommand{\DRbar}{\overline{\mathrm{DR}}}
\newcommand{\MSbar}{\overline{\mathrm{MS}}}
\newcommand{\order}{\mathcal O}
\newcommand{\ii}{\mathrm{i}}

\newcommand{\diag}{\mathop{\rm diag}}

\newcolumntype{P}[1]{>{\raggedright\arraybackslash}p{#1}}

\title{\boldmath TeV Higgsino Interpretation of the LZ High-Recoil Event with Intermediate-Scale Electroweak Gauginos}
\hypersetup{
  pdftitle={TeV Higgsino Interpretation of the LZ High-Recoil Event with Intermediate-Scale Electroweak Gauginos},
  pdfauthor={Xiaokang Du and Fei Wang}
}

\author[a,b]{Xiaokang Du}
\author[c,1]{Fei Wang\note[1]{Corresponding author.}}

\affiliation[a]{Institute of Physics, Henan Academy of Sciences, Zhengzhou
450046, P. R. China}
\affiliation[b]{Centre for Theoretical Physics, Henan Normal University,
Xinxiang 453007, P. R. China}
\affiliation[c]{School of Physics, Zhengzhou University, Zhengzhou 450000,
P. R. China}

\emailAdd{xkdu@hnas.ac.cn}
\emailAdd{feiwang@zzu.edu.cn}

\abstract{%
The high-recoil event reported by LUX--ZEPLIN motivates a TeV Higgsino with
a sub-MeV neutral-state splitting.  Away from cancellations, such a small gap
in the MSSM points to electroweak-gaugino masses of order
$10^7\,\mathrm{GeV}$, far above the Higgsino mass.  We show that the
non-universal boundary condition $M_1^G/M_2^G=-3/5$ at the $\mathrm{SU}(5)$
GUT scale cancels the leading bino and wino contributions and is preserved by
homogeneous one-loop evolution down to the Higgsino scale.  The remaining
finite-gaugino terms admit a controlled tree-level solution with a wino mass
near $120\TeV$ for a $350\keV$ gap, a $1.091\TeV$ Higgsino and
$\tan\beta=10$.  We identify mixed $\mathrm{SU}(5)$ representations that
realize the required gaugino ratio and examine the radiative sensitivity and
spectrum consistency of this construction.  Because the inelastic $Z$
coupling remains essentially unsuppressed, a full-density thermal Higgsino is
still subject to the published solar-capture bounds under their astrophysical
and transport assumptions.}

\begin{document}
\maketitle
\clearpage
\flushbottom

\section{Introduction}
\label{sec:introduction}

The LUX--ZEPLIN collaboration has extended its nuclear-recoil search to
approximately $270\keV$ and reported one event at
$E_R=248\pm23\,({\rm stat})\pm23\,({\rm sys})\keV$ with a low background
expectation~\cite{LZ:2026highrecoil}.  The profile-likelihood excess has a
maximum local significance of about $3.4\sigma$ and a global significance of
$2.6\sigma$ after the look-elsewhere effect.  Its interpretation through endothermic
scattering probes the high-velocity tail of the halo distribution
~\cite{TuckerSmith:2001inelastic,Bramante:2016inelastic,Fan:2026sea}.
Such a signal may be a hint of the supersymmetric dark matter.

 Weak-scale SUSY is one of the most promising candidates for new physics beyond the SM. It can stabilize the weak scale against quantum corrections involving energies far beyond the weak scale, realize successful gauge coupling unification, and provide viable dark matter candidates. A nearly pure MSSM Higgsino is a interesting thermal WIMP candidate with mass near
$1.1\TeV$.  In the limit of decoupled electroweak gauginos, the two neutral
Higgsino Weyl fields form a pseudo-Dirac pair. Majorana bino and wino masses
break Higgsino number through dimension-five operators and generate the
neutral-state splitting. Light Higgsino dark matter in high-scale SUSY were analyzed in ref.~\cite{Nagata:2014wma}.

The LZ-motivated
analyses find a representative mass near $1.091\TeV$ and a splitting of a few
hundred keV~\cite{Su:2026inelastic,Freese:2026higgsino,Wu:2026higgsino}. Recoil-level Higgsino studies favor
reference splittings around $0.34$--$0.37\MeV$~\cite{DiMauro:2026kinematic,Freese:2026higgsino,Wu:2026higgsino}.
A solar-capture analysis of a full-density thermal Higgsino instead obtains
$\dmzero\gtrsim506\keV$ from inelastic slowing and
$\dmzero\gtrsim566\keV$ after loop-induced elastic slowing
~\cite{Pospelov:2026solar}.  
Alternative inelastic electroweak-doublet and
dark-photon interpretations are discussed in refs.~\cite{Visinelli:2026inelastic,Yamashita:2026darkphoton},
while absorption has also been considered~\cite{Lou:2026absorption}.

The explanation of LZ signal with light Higgsino dark matter and relatively heavy (other) SUSY particles calls for a top-down explanation---for instance, from a typical SUSY-breaking mechanism. The separation of the Higgsino and supersymmetry-breaking scales had also been studied in gravity and gauge mediation
~\cite{Evans:2014pka,Giudice:1998bp}.  The $\mu$ and $B_\mu$ mechanisms
relevant to the LZ interpretation are discussed in ref.~\cite{Fan:2026sea}. The associated Higgs-sector conditions are independent of the gaugino boundary cancellation condition in this work.

We would like to discuss the UV origin of the opposite-sign cancellation for neutral-state splitting.
The ratio between gaugino masses and the corresponding gauge couplings
are constant under one-loop renormalization group running up to
small (and known) two-loop effects and possibly threshold effects
near unification scale. Since the gauge couplings are observed to
unify at the unification scale, it is a popular assumption that the
gaugino masses also unify near that scale. When additional higher
dimensional operators are included, the gauge kinetic
terms will get additional contributions after GUT gauge symmetry
breaking so that they will no longer unify at the Grand Unification
scale. Besides, with proper F-term VEV, gaugino masses will no
longer be universal at the GUT scale~\cite{Ellis:1985,Drees:1985,Martin:2009ad,Chakrabortty:2009}.

 Under proper $\mathrm{SU}(5)$ normalization, the leading cancellation
requires the target ratio $M_1^G/M_2^G=-3/5$.  At this fixed ratio, the first surviving term scales
as $\mu\sin2\beta/M_2^2$ and selects a gaugino scale of order $10^2\TeV$
for a sub-MeV gap. Such a low scale gaugino scenario differs from the inverse-mass scaling of the
same-sign decoupling branch, whose scale is near $10^7\GeV$ and has a
concrete high-scale realization in ref.~\cite{Yin:2026highrecoil}. We determine the tree-level roots on this boundary, classify its
mixed-representation realizations, and calculate the local sensitivity to
the bino--wino alignment.

Section~\ref{sec:infrared} derives the neutral-Higgsino splitting and the
heavy-gaugino expansion.  Section~\ref{sec:cancellation} gives the unified
boundary conditions and their tree-level solutions.  The phenomenological
reference scales and solar-capture constraint are discussed in
Section~\ref{sec:phenomenology}.  Section~\ref{sec:consistency} treats the
boundary sensitivity, target inversion and spectrum consistency, followed
by the conclusions in Section~\ref{sec:conclusions}.   

\section{Neutral-Higgsino splitting and the cancellation branch}
\label{sec:infrared}

We work with real parameters, choose the wino phase so that $M_2>0$, and
restrict our numerical study to $\mu>0$.  The $M_a$ are signed Majorana
mass parameters, whereas physical masses are nonnegative; in particular,
$M_1<0$ on the cancellation branch.  We define $M_a^G\equiv M_a(M_G)$,
$\tan\beta=v_u/v_d$, $s_\beta=\sin\beta$, and $c_\beta=\cos\beta$.
The reduced Planck scale is $\MPl=2.435\times10^{18}\GeV$. The hypercharge coupling is normalized according to SU(5) GUT as $g_1=\sqrt{5/3}\,g_Y$.

\subsection{Neutralino matrix and finite-gaugino expansion}
\label{subsec:neutralino-matrix}

For the two-component fields
$\psi^0=(\widetilde B,\widetilde W^0,\widetilde H_d^0,\widetilde H_u^0)^T$,
our mass convention is
$\mathcal L_{\rm mass}=-\tfrac12\psi^{0T}\mathcal M_N\psi^0+\mathrm{h.c.}$,
with the tree-level matrix
\begin{equation}
 {\cal M}_N=
 \begin{pmatrix}
 M_1&0&-m_Zs_Wc_\beta&m_Zs_Ws_\beta\\
 0&M_2&m_Zc_Wc_\beta&-m_Zc_Ws_\beta\\
 -m_Zs_Wc_\beta&m_Zc_Wc_\beta&0&-\mu\\
 m_Zs_Ws_\beta&-m_Zc_Ws_\beta&-\mu&0
 \end{pmatrix}.
 \label{eq:neutralino-matrix}
\end{equation}
In a split-spectrum calculation, the off-diagonal entries are instead
specified by four independently running Higgs--Higgsino--gaugino couplings,
\begin{equation}
 {\cal M}^{\rm split}_N(Q)=
 \begin{pmatrix}
 M_1&0&-\tfrac12\widetilde g_{Yd}v&\tfrac12\widetilde g_{Yu}v\\
 0&M_2&\tfrac12\widetilde g_{2d}v&-\tfrac12\widetilde g_{2u}v\\
 -\tfrac12\widetilde g_{Yd}v&\tfrac12\widetilde g_{2d}v&0&-\mu\\
 \tfrac12\widetilde g_{Yu}v&-\tfrac12\widetilde g_{2u}v&-\mu&0
 \end{pmatrix}_{Q} .
 \label{eq:split-neutralino-matrix}
\end{equation}
At a supersymmetric matching threshold,
$\widetilde g_{Yu}=g_Y\sin\beta$,
$\widetilde g_{Yd}=g_Y\cos\beta$,
$\widetilde g_{2u}=g_2\sin\beta$, and
$\widetilde g_{2d}=g_2\cos\beta$, with $g_Y=\sqrt{3/5}\,g_1$.
They separate under running below heavy thresholds~\cite{Nagata:2014wma}.
All numerical results below use the common-scale convention of
eq.~\eqref{eq:neutralino-matrix}; the split form specifies the couplings
needed for threshold matching.

To second order in the off-diagonal electroweak mixing, namely at
$\order(m_Z^2)$, but without expanding in $\mu/M_i$, the
neutral-Higgsino splitting is approximated by
\begin{equation}
 \delta m_0^{(2)}\equiv m_Z^2\left|
 s_W^2\frac{M_1+\mu\sin2\beta}{M_1^2-\mu^2}
 +c_W^2\frac{M_2+\mu\sin2\beta}{M_2^2-\mu^2}\right| .
 \label{eq:finite-gaugino-splitting}
\end{equation}
Writing $d=\min_{i=1,2;\,\pm}|M_i\pm\mu|$, perturbation theory gives
\begin{equation}
 \delta m_0^{\rm tree}=\delta m_0^{(2)}
 +\order\!\left(\frac{m_Z^4}{|\mu|d^2},\frac{m_Z^4}{d^3}\right).
 \label{eq:finite-gaugino-remainder}
\end{equation}
This absolute error estimate requires $d\gg m_Z$ and
$m_Z^2/d\ll|\mu|$ and includes mixed bino--wino terms.
It need not be small relative to a tuned residual gap.
Equation~\eqref{eq:finite-gaugino-splitting} displays the cancellation
between the two sources~\cite{Martin:2024curtain} and all numerical target
roots below use exact diagonalization.

For $|M_{1,2}|\gg |\mu|,m_Z$, the Schur complement of the gaugino block gives
\begin{align}
 {\cal M}_{\widetilde H}^{\rm eff}
 &=\begin{pmatrix}0&-\mu\\-\mu&0\end{pmatrix}
 -\kappa\begin{pmatrix}
 c_\beta^2&-s_\beta c_\beta\\
 -s_\beta c_\beta&s_\beta^2
 \end{pmatrix},
 \label{eq:higgsino-schur}\\
 \kappa&\equiv m_Z^2\left(\frac{s_W^2}{M_1}+
                       \frac{c_W^2}{M_2}\right).
 \label{eq:kappa-definition}
\end{align}
Equation~\eqref{eq:higgsino-schur} is the leading broken-phase
dimension-five matching.  Its diagonal Majorana entries arise from two
Higgsino-number-violating operators,
\begin{equation}
 {\cal L}_{5}^{\rm HNV}=c_1\mathcal O_1+c_2\mathcal O_2+\mathrm{h.c.},
 \qquad
 \mathcal O_1=(H^\dagger\widetilde H_u)^2,
 \qquad
 \mathcal O_2=(H\mathbin{\cdot}\widetilde H_d)^2 .
 \label{eq:eft-dim5-operators}
\end{equation}
Here $H\mathbin{\cdot}\widetilde H_d
\equiv\epsilon_{ij}H^i\widetilde H_d^j$, with $\epsilon_{12}=+1$;
each square contracts the two Weyl-spinor indices antisymmetrically.
The complete EFT also contains two Higgsino-number-preserving operators.
They contribute to the off-diagonal entry in
eq.~\eqref{eq:higgsino-schur} and to the charged--neutral splitting.
Thus $\kappa$ is not
a Wilson coefficient by itself.  It is the CP-conserving broken-phase
combination of $c_1$ and $c_2$ that controls the leading neutral--neutral
gap~\cite{Nagata:2014wma}.

Using non-degenerate perturbation theory on the two unperturbed eigenvectors,
we find for $\mu>0$ and $\kappa>0$
\begin{align}
 m_{\chi^0_1}^{\rm tree}&=\mu-\frac{\kappa}{2}(1+\sin2\beta)
 +\order\!\left(\frac{m_Z^2\mu}{M_{1,2}^2},\frac{m_Z^4}{M_{1,2}^3},\frac{\kappa^2}{|\mu|}\right),
 \label{eq:chi-one-mass}\\
 m_{\chi^0_2}^{\rm tree}&=\mu+\frac{\kappa}{2}(1-\sin2\beta)
 +\order\!\left(\frac{m_Z^2\mu}{M_{1,2}^2},\frac{m_Z^4}{M_{1,2}^3},\frac{\kappa^2}{|\mu|}\right),
 \label{eq:chi-two-mass}\\
 \delta m_0^{\rm tree}&\equiv m_{\chi^0_2}^{\rm tree}-m_{\chi^0_1}^{\rm tree}
 =|\kappa|+\order\!\left(\frac{m_Z^2|\mu|}{M_{1,2}^2},
                            \frac{m_Z^4}{M_{1,2}^3},\frac{\kappa^2}{|\mu|}\right).
 \label{eq:neutral-splitting-leading}
\end{align}
The individual masses depend on $\tan\beta$, but their leading difference
does not.  For negative $\kappa$, the ordering of the two Majorana states is exchanged.
The physical splitting remains $|\kappa|$ at this order.  The charged-neutral
mass difference is a separate quantity, dominated in the pure-Higgsino limit
by electroweak loops and of order $350\MeV$, not $350\keV$.

For later use we distinguish the signed ordering variable from the physical gap,
\begin{equation}
 \Delta_s^{\rm tree}\equiv|z_-|-|z_+|,
 \qquad
 \delta m_0^{\rm tree}=|\Delta_s^{\rm tree}|,
 \label{eq:signed-definition}
\end{equation}
where $z_+$ and $z_-$ are the signed eigenvalues continuously
connected to $+\mu$ and $-\mu$ as $m_Z\to0$.  A zero of $\Delta_s$ is a
physical mass degeneracy and an exchange of the mass ordering.

\subsection{Phenomenological reference splittings and the two branches}
\label{subsec:two-branches}

Table~\ref{tab:reference-splittings} collects reference splittings from
distinct phenomenological analyses.
\begin{table}[!ht]
 \centering
 \small
 \begin{tabular}{@{}lll@{}}
 \toprule
 Reference & Interpretation & Use here \\
 \midrule
 $340$--$360\keV$ & representative LZ Higgsino interval & recoil reference \\
 $350\keV$ & benchmark of this paper & normalization point \\
 $371\keV$ & recoil-level Higgsino benchmark & updated LZ reference \\
 $506\keV$ & tree-only solar-capture reference & literature scale \\
 $566\keV$ & loop-informed solar-capture reference & literature scale \\
 \bottomrule
 \end{tabular}
 \caption{Phenomenological reference splittings from
 refs.~\cite{Fan:2026sea,Freese:2026higgsino,Wu:2026higgsino,DiMauro:2026kinematic,Pospelov:2026solar}.
 The terrestrial and solar values are not statistically combined into a
 likelihood or confidence interval.}
 \label{tab:reference-splittings}
\end{table}

When the leading dimension-five term dominates, the illustrative gap
$\delta m_0=350\keV$ corresponds to
\begin{equation}
 \left|\frac{s_W^2}{M_1}+\frac{c_W^2}{M_2}\right|
 \simeq4.21\times10^{-8}\GeV^{-1},
 \qquad
 \frac{m_Z^2}{\delta m_0}\simeq2.38\times10^7\GeV.
 \label{eq:350kev-wilson-coefficient}
\end{equation}
The second number characterizes the common-sign decoupling scale.  

For common signs, the decoupling solution is obtained directly.  Two useful
normalizations are
\begin{align}
 M_1=M_2&\simeq2.38\times10^7\GeV,
 \label{eq:equal-gaugino}\\
 M_1=\rho_{12}(Q_{\rm ref})M_2, &\qquad
 (M_1,M_2)(Q_{\rm ref})\simeq(1.54,2.88)\times10^7\GeV.
 \label{eq:gmsb-gaugino}
\end{align}
The second relation is the one-loop low-energy pattern from a unified or
minimal gauge-mediated boundary condition, evaluated using
$M_2=m_Z^2(s_W^2/\rho_{12}+c_W^2)/\delta m_0$ at $Q_{\rm ref}$.
A concrete common-sign high-scale realization is given in
ref.~\cite{Yin:2026highrecoil}.

There is also an opposite-sign branch,
\begin{equation}
 \frac{M_1}{M_2}=-\tan^2\theta_W(Q_{\rm ref})
 \label{eq:cancellation-leading}
\end{equation}
at which the leading dimension-five coefficient vanishes~\cite{Chun:2016degenerate,Martin:2024curtain}.
A finite splitting on this branch has two distinct realizations.  On the
fixed-ratio branch, $M_1/M_2$ is imposed by the GUT boundary and the finite
$\mu/M_i$ terms determine the overall scale.  Alternatively, at fixed $M_2$
one may retune the ratio away from eq.~\eqref{eq:cancellation-leading}.
The latter construction is given in Appendix~\ref{app:retuning}.

\section{Cancellation branch from non-universal unified gaugino masses}
\label{sec:cancellation}

The cancellation mechanism concerns the relative electroweakino boundary
condition.  The origin of the TeV-scale $\mu$ parameter and the associated
$B_\mu$ and electroweak-symmetry-breaking conditions remain independent
model-building requirements.

\subsection{Canonical \texorpdfstring{$\mathrm{SU}(5)$}{SU(5)} normalization and the GUT target}
\label{subsec:su5}
Universal gaugino masses can arise from the minimal forms of the gauge kinetic functions. 
For non-minimal form of Kahler potential, more general superpotential terms and non-minimal gauge kinetic
functions, the soft masses, trilinear couplings and gaugino masses
are in general non-universal. The non-minimal gauge kinetic terms can be generated by higher dimensional operators with GUT breaking effects.
\begin{equation}
W \supset \frac12 \mathrm{Tr}\left[ W^a W^b \left( \tau \delta_{ab} +c_{\bf r} \frac{\Phi_{{\bf r},ab}}{M_*} \right) \right]\,,
\end{equation}
with $\Phi_{{\bf r},ab}$ the representation ${\bf r}$ Higgs and  $c_{\bf r}$ the coefficient for the higher dimensional operators.
After substituting the VEV of representation ${\bf r}$ Higgs of the general form 
\beqa
\Phi_{{\bf r},ab}=(v_{\bf r}+\theta^2 F_{\bf r}) M_{{\bf r};ab}~,
\eeqa
 soft SUSY breaking gaugino masses can be obtained (and also the GUT relation for the gauge couplings will be modified). 
Here $M_{{\bf r};ab}$ is the relevant GUT group factor for the VEV that are consistent with the breaking of the GUT group into the Standard Model. When a nonsinglet field has both a scalar expectation value and an
$F$-term in the gauge kinetic function, canonical normalization must be carried
out before identifying the gaugino mass ratios.

For $SU(5)$ GUT, the matter contents include 
\beqa 
{\bf{\bar{ 5}}}=(~D_i^c,~L^i)~,~~{\bf
10}=(~Q_i,~U^i_c,~E^c_i)~. 
\eeqa

The non-universal gaugino masses arise from the $\mathrm{SU}(5)$ decomposition
\begin{equation}
 (\mathbf{24}\otimes\mathbf{24})_{\rm S}=
 \mathbf1\oplus\mathbf{24}\oplus\mathbf{75}\oplus\mathbf{200},
 \label{eq:su5-product}
\end{equation}

The group structure for the VEV of ${\bf 24}$ dimensional representation adjoint Higgs $\Phi$ can be written as
\beqa
<\Phi_{\bf 24}>\sim \sqrt{\f{3}{5}}\(\bea{ccccc}\f{1}{3}&&&&\\&\f{1}{3}&&&\\&&\f{1}{3}&&\\&&&-\f{1}{2}&\\&&&&-\f{1}{2}\eea\)~.
\eeqa
with the $5\tm 5$ matrix normalization factor $(c = 1/2)$. 

The group structure for the VEV of ${\bf 75}$
dimensional representation Higgs can be written as
 \beqa
<\Phi_{\bf 75}>^{ik}_{jl}\sim \f{1}{\sqrt{12}}\[\Delta_{cj}^{[i}\Delta_{kl}^{k]}+2\Delta_{wj}^{[i}\Delta_{wl}^{k]}
-\f{1}{2}\delta^{[i}_{j}\delta^{k]}_{l}\]~,
 \eeqa
with
\beqa
\Delta_c=diag(~1,~1,~1,~0,~0)~,\Delta_w=diag(~0,~0,~0,~1,~1)~.
\eeqa
Here we normalize the VEV according to the $10\tm 10$ matrix normalization factor $(c = 3/2)$.
The group structure for the VEV can also be written as $10\tm 10$ matrix
\beqa 
<\Phi_{\bf 75}>\sim \f{1}{\sqrt{12}}
diag(\underbrace{~1,\cdots,~1}_3,\underbrace{-1,\cdots,-1}_{6},3)~,
\eeqa 
which is orthogonal to representations for ${\bf 24}$ dimensional Higgs VEV.

Similarly, the group structure for the VEV of ${\bf 200}$ dimensional representation Higgs can be written as a $(15 \times 15)$
traceless diagonal matrix with normalization factor $(c = 7/2)$
\begin{equation}
< \Phi_{\bf 200} > \sim \frac{1}{\sqrt{12}} \, \mathrm{diag}(1,1,1,1,1,1,-2,-2,-2,-2,-2,-2,2,2,2) .
\end{equation}

At the unification scale, the GUT-normalized coupling satisfies
$$g_1(M_G)=g_2(M_G)=g_3(M_G)=g_G.$$ 
From the hypercharge assignment of the matter contents that fitted into SU(5) representations, the normalization $g_Y^2=3g_1^2/5$ is adopted.  Therefore
\begin{equation}
 \sin^2\theta_W(M_G)=\frac38,
 \qquad
 \tan^2\theta_W(M_G)=\frac35 .
 \label{eq:su5-weak-angle}
\end{equation}
The one-loop MSSM RGE evolution predict that 
$M_a(Q)/g_a^2(Q)=M_a^G/g_G^2$ for $a=1,2,3$ because $M_a/g_a^2$ is a one-loop RGE invariant.

 The leading common-scale
cancellation condition is equivalent to
\begin{align}
 \frac{g_Y^2(Q)}{M_1(Q)}+\frac{g_2^2(Q)}{M_2(Q)}
 &=g_G^2\left(\frac{3}{5M_1^G}+\frac{1}{M_2^G}\right)=0,\label{eq:gut-cancellation-map}\\\Rightarrow
 \frac{M_1^G}{M_2^G}&=-\frac35 .\label{eq:GUT-target}
\end{align}
Thus the GUT relation guarantee the leading cancellation, while the residual
finite-$\mu/M_i$ terms determine the overall electroweakino scale.
Requiring the cancellation selects this ratio, which is not required by gauge coupling unification alone. GUT thresholds, non-canonical gauge kinetic terms, and other high-order Planck-suppressed 
operators can shift the relation in eq.~\eqref{eq:GUT-target}.

The new contributions to gaugino masses for the Higgs fields in various representations are given by
~\cite{Ellis:1985,Drees:1985,Martin:2009ad,Chakrabortty:2009}
\begin{equation}
 \begingroup
 \renewcommand{\arraystretch}{1.65}
 \setlength{\arraycolsep}{3.5pt}
 \resizebox{0.86\textwidth}{!}{$
 \begin{array}{@{}ccccc@{}}
  \toprule
  R & \mathbf1 & \mathbf{24} & \mathbf{75} & \mathbf{200} \\
  \midrule
  (M_1,M_2,M_3)_R
  & (1,1,1)
  & \left(\frac{1}{\sqrt{15}},\frac{3}{\sqrt{15}},-\frac{2}{\sqrt{15}}\right)
  & \left(\frac{4}{\sqrt3},-\frac{12}{5\sqrt3},-\frac{4}{5\sqrt3}\right)
  & \left(\frac{1}{\sqrt{21}},\frac{1}{5\sqrt{21}},\frac{1}{10\sqrt{21}}\right) \\
  (C_1^{(R)},C_2^{(R)},C_3^{(R)})
  & (1,1,1)
  & (-\tfrac12,-\tfrac32,1)
  & (-5,3,1)
  & (10,2,1) \\
  \bottomrule
 \end{array}
 $}
 \endgroup
 \label{eq:su5-vectors}
\end{equation}
with the second row the normalized ratio vectors after redefining the relevant coefficients.

Let $m_R$ denote the dimension-one coefficient multiplying each dimensionless ratio vectors.
The notation $\mathbf1+\mathbf{75}$ denotes two independent contributions,
not a single irreducible representation.  Choosing
$m_{\mathbf{75}}/m_{\mathbf1}=1/2$ gives the boundary vector
\begin{equation}
 (M_1,M_2,M_3)_G
 =m_{\mathbf1}\left[(1,1,1)+\frac12(-5,3,1)\right]
 =\frac{m_{\mathbf1}}2(-3,5,3),
 \label{eq:canonical-one-75}
\end{equation}
whose one-loop reference evolution gives
\begin{equation}
 r_{1+75}(Q)\equiv\frac{M_1}{M_2}(Q)
 =-\frac35\rho_{12}(Q),
 \qquad \rho_{12}(Q)\equiv\frac{g_1^2(Q)}{g_2^2(Q)} .
 \label{eq:canonical-one-75-ratio}
\end{equation}
With $g_1(M_G)=g_2(M_G)$ and the illustrative $Q_{\rm ref}=1\TeV$ couplings in
eq.~\eqref{eq:one-tev-couplings}, one obtains
$r_{1+75}(Q_{\rm ref})=-\tan^2\theta_W(Q_{\rm ref})\simeq-0.32198255$.

Using the exact neutralino characteristic equation in
Appendix~\ref{app:quartic} with $\mu=1.091\TeV$,
$\tan\beta=10$, and $M_2(Q_{\rm ref})=100\TeV$ gives
\begin{equation}
 \delta m_0^{\rm tree}\big|_{1+75}\simeq494\keV .
 \label{eq:canonical-one-75-gap}
\end{equation}
At this fixed ratio, the GUT group strucutre fixes the relative
sign and ratio, while the overall normalization is determined by the
tree-level target,
\begin{equation}
 \delta m_0^{\rm tree}\!\left(M_2(Q_{\rm ref})\right)=350\keV
 \quad\Longrightarrow\quad
 (M_1,M_2)(Q_{\rm ref})\simeq(-38.649,\,120.035)\TeV .
 \label{eq:canonical-one-75-target}
\end{equation}

\subsection{Fixed-ratio residual splitting and the high-mass branch}
\label{subsec:exact-cancellation}

The fixed-ratio relation removes the leading dimension-five coefficient while
the finite-$\mu/M_i$ terms determine the residual gap.  Let $z$
denote a signed neutralino eigenvalue and define
the energy-dependent Schur coefficient
\begin{equation}
 \kappa(z)=m_Z^2\left(\frac{s_W^2}{M_1-z}
 +\frac{c_W^2}{M_2-z}\right).
 \label{eq:energy-dependent-kappa}
\end{equation}
The coefficient in eq.~\eqref{eq:kappa-definition} is $\kappa=\kappa(0)$.
From the exact tree-level characteristic polynomial given in
Appendix~\ref{app:quartic} and factoring out the $(z-M_1)$ and $(z-M_2)$ factor, the eigenvalues of the two Higgsino-like states are the roots of 
\begin{equation}
 z^2-\mu^2+\kappa(z)\bigl(z+\mu\sin2\beta\bigr)=0.
 \label{eq:energy-dependent-root}
\end{equation}
which is very near $z=\pm\mu$. No expansion in $\mu/M_{1,2}$ is made in our
numerical solution.

For analytic estimations, we expand $\kappa(z)$ 
\begin{equation}
\kappa(z)=m_Z^2\sum_{n\geq0}z^n A_n~, \qquad A_n\equiv\frac{s_W^2}{M_1^{n+1}}+\frac{c_W^2}{M_2^{n+1}}.
\label{eq:An-expansion}
\end{equation}
The coefficients $A_n$ have mass dimension $-n-1$.
Convergence requires $|z|<\min(|M_1|,|M_2|)$.
Near the Higgsino roots this becomes $|\mu|<\min(|M_1|,|M_2|)$.
We call the expansion controlled when $|\mu|/\min(|M_1|,|M_2|)<0.1$.
Expanding the two roots around $\pm\mu$ gives the physical splitting
\begin{equation}
 \delta m_0^{\rm tree}=\left|m_Z^2\left[A_0+\mu\sin2\beta\,A_1
 +\mu^2A_2+\mu^3\sin2\beta\,A_3+\cdots\right]
 +\mathcal R_{\rm EW}^{(4+)}\right| .
 \label{eq:cancellation-expanded-gap}
\end{equation}
Here $\mathcal R_{\rm EW}^{(4+)}$ has mass dimension one and denotes fourth
and higher orders in electroweak mixing, with the parametric estimate in
eq.~\eqref{eq:finite-gaugino-remainder}.  It need not vanish when
$\kappa(0)=0$.  

The first term is the familiar dimension-five coefficient. When $A_0$ vanishes, that is,
$r_{\rm can}=-s_W^2/c_W^2$ for $r\equiv M_1/M_2$, the next terms are fixed by the same two
mass parameters.  Defining $\overline A_n\equiv A_n|_{r=r_{\rm can}}$ gives
\begin{equation}
 \overline A_1=\frac{c_W^2}{s_W^2M_2^2},
 \qquad
 \overline A_2=\frac{c_W^2(s_W^2-c_W^2)}{s_W^4M_2^3} .
 \label{eq:cancellation-An-star}
\end{equation}

On the fixed canonical ratio, the first surviving term gives
\begin{equation}
 \delta m_{0,\rm lead}^{\rm can}\simeq
 \frac{m_Z^2|\mu\sin2\beta|}{M_2^2}\frac{c_W^2}{s_W^2},\\
 \end{equation}
 which can fix (to leading order) the low energy scale of $M_2$ 
 \begin{equation}
 M_2\simeq
 \left[\frac{m_Z^2|\mu\sin2\beta|}{\delta m_0^{\rm tree}}
 \frac{c_W^2}{s_W^2}\right]^{1/2} .
 \label{eq:fixed-ratio-scale}
\end{equation}

 With $\mu=1.091\TeV$ and $\tan\beta=10$, the $\overline A_1$ term
alone gives $M_2^{(0)}\simeq126.256\TeV$.  Including $\overline A_2$ gives
$119.998\TeV$, and including $\overline A_3$ gives $120.041\TeV$, compared with
the exact root $120.035\TeV$.  The convergence of the numerical $M_2$ value indicates that the heavy-gaugino
expansion is consistent on the asymptotic branch, where
$|\mu|/\min(|M_1|,|M_2|)\simeq0.028$.

At the exact root, the successive contributions are
\begin{equation}
 \begin{gathered}
 m_Z^2\overline A_0=0,\qquad
 m_Z^2\mu\sin2\beta \overline A_1\simeq387.2\keV,\\
 m_Z^2\mu^2\overline A_2\simeq-37.4\keV,\qquad
 m_Z^2\mu^3\sin2\beta \overline A_3\simeq0.24\keV .
 \end{gathered}
 \label{eq:canonical-residual-terms}
\end{equation}
The first nonzero term therefore fixes the electroweakino scale, while the
$\overline A_2$ term supplies the ten-percent correction needed for the $350\keV$
solution.

\begin{table}[!ht]
 \centering
 \small
 \begin{tabular}{@{}cccc@{}}
 \toprule
 $\delta m_0$ & $M_2^{(0)}$ from $\overline A_1$ & $M_2$ exact high branch & $M_1$ exact high branch \\
 \midrule
 $300\keV$ & $136.4\TeV$ & $130.187\TeV$ & $-41.918\TeV$ \\
 $324\keV$ & $131.2\TeV$ & $125.022\TeV$ & $-40.255\TeV$ \\
 $350\keV$ & $126.3\TeV$ & $120.035\TeV$ & $-38.649\TeV$ \\
 $500\keV$ & $105.6\TeV$ & $99.315\TeV$ & $-31.978\TeV$ \\
 \bottomrule
 \end{tabular}
 \caption{Leading and exact high-branch scales on the canonical cancellation
 line for $\mu=1.091\TeV$ and $\tan\beta=10$.  The exact columns are
 displayed to $1\GeV$ and denote signed Majorana reference parameters,
 not pole masses.  A negative entry fixes a relative phase, not a negative
 physical mass.}
 \label{tab:scale-range}
\end{table}

\subsection{Higgsino-dominated tree-level roots at the canonical ratio}
\label{subsec:root-structure}

At the canonical ratio, the exact tree-level spectrum gives
three values of $M_2$ with a Higgsino-like neutral pair and a
$350\keV$ splitting,
\begin{equation}
 M_2(Q_{\rm ref})\simeq11.818\TeV,\qquad 12.007\TeV,\qquad 120.035\TeV .
 \label{eq:three-canonical-roots}
\end{equation}
The first two lie on opposite sides of the zero of $\Delta_s^{\rm tree}$,
which occurs at $M_2(Q_{\rm ref})\simeq11.910\TeV$; the third is the asymptotic
finite-$\mu/M_i$ branch considered here.

\begin{figure}[!htbp]
 \centering
 \includegraphics[width=0.82\textwidth]{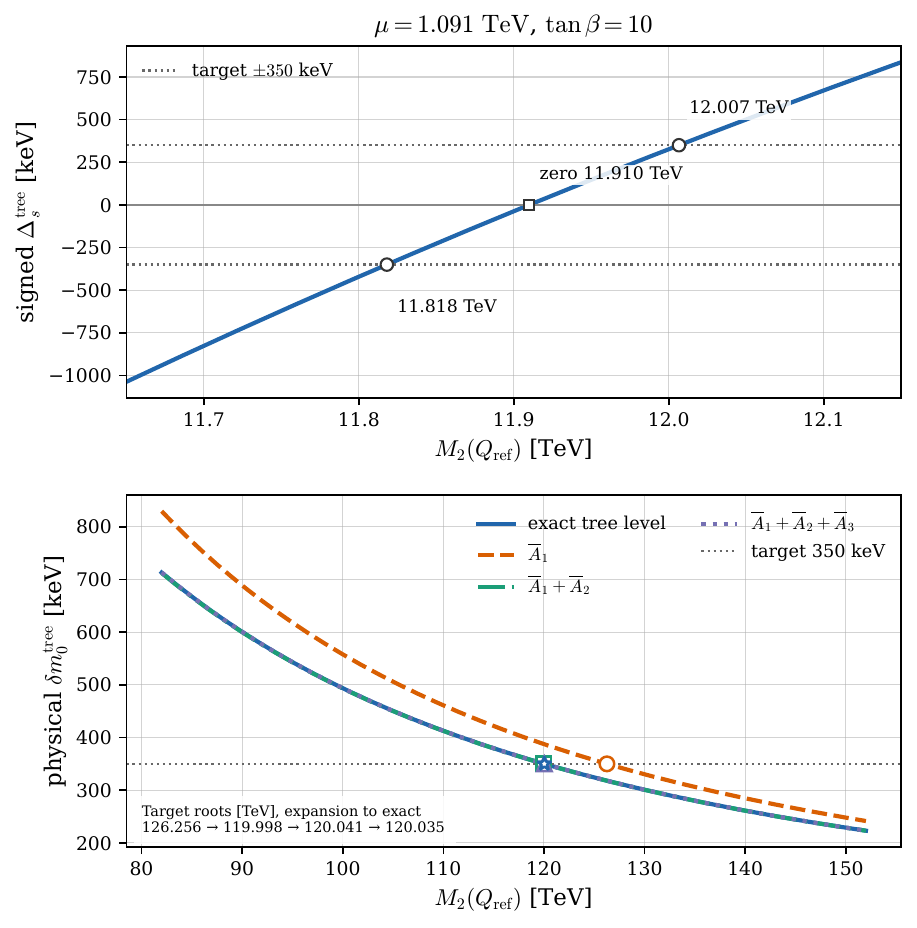}
\caption{Tree-level neutral-Higgsino splitting on the fixed-ratio line for
 $M_1/M_2=r_{\rm can}\simeq-0.32198255$, $\mu=1.091\TeV$, and $\tan\beta=10$.  The upper panel shows
 $\Delta_s^{\rm tree}=|z_-|-|z_+|$ near the two low-scale roots
 at $11.818\TeV$ and $12.007\TeV$ and the intervening physical-mass
 degeneracy.  The lower panel compares the exact high branch with the
 $\overline A_1$, $\overline A_1+\overline A_2$, and $\overline A_1+\overline A_2+\overline A_3$ expansions.  The convergence
 chain ends at the $120.035\TeV$ high-mass root.  All quantities are
 matching-scale tree-level results.}
 \label{fig:canonical-gap-roots}
\end{figure}

A separate level-crossing root
near $M_2\simeq3.38\TeV$ is not counted because the nominal Higgsino pair is
then strongly mixed with a gaugino and the expansion about $\pm\mu$ is not a
controlled description.  The asymptotic branch satisfies
$|\mu|/\min(|M_1|,|M_2|)<0.1$, whereas the low-scale pair does not.
The latter still satisfies the geometric convergence condition, but requires
more terms than the controlled high-branch approximation used here.
The signed crossing and asymptotic falloff are displayed in
Fig.~\ref{fig:canonical-gap-roots}.

\subsection{General mixed \texorpdfstring{$\mathrm{SU}(5)$}{SU(5)} spurions}
\label{subsec:representation-rge}

For the representations in eq.~\eqref{eq:su5-product}, the effective gauge
kinetic function is
\beqa
 f_{AB}&=&f_0\delta_{AB}+
 \zeta_{\mathbf1}\frac{<\Phi_{\mathbf1}>}{M_{\rm cut}}\delta_{AB}+
 \zeta_{\mathbf{24}}\frac{<\Phi_{\mathbf{24}}>_{AB}}{M_{\rm cut}}+\zeta_{\mathbf{75}}\frac{<\Phi_{\mathbf{75}}>_{AB}}{M_{\rm cut}}\\
 &+&\zeta_{\mathbf{200}}\frac{<\Phi_{\mathbf{200}}>_{AB}}{M_{\rm cut}}+\cdots .
 \label{eq:general-gauge-kinetic}
\eeqa
Here $\zeta_{\mathbf r}$ denotes the coefficients for the non-renormalizable operators, $A,B$ are adjoint indices and $M_{\rm cut}$ is the suppression scale of the non-renormalizable gauge-kinetic operators, not necessarily $\MPl$.  Each $<\Phi_R>$ denotes the VEV of various high representation Higgs fields, which lies along the
SM-singlet direction of its representation. As noted before, the VEV can be factored out a group strucutre tensors $T_R$  that act on the
three SM gauge blocks with coefficients $C_a^{(R)}$ in
eq.~\eqref{eq:su5-vectors}.

With $\frac14\int d^2\theta\,f_{AB}W^AW^B+\mathrm{h.c.}$,
$\operatorname{Re}f_0=g_G^{-2}$ and negligible non-universal scalar
contributions, canonical normalization gives
\begin{equation}
 m_R=\frac{g_G^2\zeta_RF_R}{2M_{\rm cut}},
 \qquad R=\mathbf1,\mathbf{24},\mathbf{75},\mathbf{200},\cdots
 \label{eq:spurion-mass-normalization}
\end{equation}
We use $<\Phi_R>|_{\theta^2}=F_R$ and
$W_\alpha^A|_{\theta=0}=-\ii\lambda_{{\rm g},\alpha}^A$; after canonical
normalization the gaugino mass term has the sign specified in
Section~\ref{subsec:neutralino-matrix}.

\Needspace{8\baselineskip}
The four spurions are independent and their $F$ terms are constrained only
by the imposed cancellation. With proper normalization and redefinition of the coefficients, the gaugino masses at the unification scale are then given by
\begin{equation}
 \begin{gathered}
 M_1^G=m_{\mathbf1}-\frac12m_{\mathbf{24}}-5m_{\mathbf{75}}+10m_{\mathbf{200}},\\
 M_2^G=m_{\mathbf1}-\frac32m_{\mathbf{24}}+3m_{\mathbf{75}}+2m_{\mathbf{200}},\\
 M_3^G=m_{\mathbf1}+m_{\mathbf{24}}+m_{\mathbf{75}}+m_{\mathbf{200}} .
 \end{gathered}
 \label{eq:general-gaugino-masses}
\end{equation}
The target relation $M_1^G/M_2^G=-3/5$ defines the plane
\begin{equation}
 8m_{\mathbf1}-7m_{\mathbf{24}}-16m_{\mathbf{75}}+56m_{\mathbf{200}}=0 .
 \label{eq:cancellation-plane}
\end{equation}
This constraint admits several representation choices.
Table~\ref{tab:mixed-su5-boundaries} gives three minimal two-component
boundaries with different gluino ratios
\begin{table}[!h]
 \centering
 \small
 \begin{tabular}{@{}ccc@{}}
 \toprule
 combination & coefficient ratio & $M_1^G:M_2^G:M_3^G$ \\
 \midrule
 $\mathbf1+\mathbf{24}$ & $m_{\mathbf{24}}/m_{\mathbf1}=8/7$ & $-3/5:1:-3$ \\
 $\mathbf1+\mathbf{75}$ & $m_{\mathbf{75}}/m_{\mathbf1}=1/2$ & $-3/5:1:3/5$ \\
 $\mathbf1+\mathbf{200}$ & $m_{\mathbf{200}}/m_{\mathbf1}=-1/7$ & $-3/5:1:6/5$ \\
 \bottomrule
 \end{tabular}
 \caption{Minimal mixed $\mathrm{SU}(5)$ boundaries that impose the ideal leading
 cancellation.  The entries are signed mass ratios.
 The different $M_3^G$ entries affect two-loop running and
 colored thresholds.}
\label{tab:mixed-su5-boundaries}
\end{table}

\subsection{An illustrative \texorpdfstring{$\mathbf1+\mathbf{75}+\mathbf{200}$}{1+75+200} boundary}
\label{subsec:one-75-200}

The cancellation plane also admits three-direction solutions.  For
example, taking $m_{\mathbf{24}}=0$ and $m_{\mathbf{200}}=0.05m_{\mathbf1}$ gives
$m_{\mathbf{75}}=0.675m_{\mathbf1}$ from eq.~\eqref{eq:cancellation-plane}, and hence
\begin{equation}
 (M_1^G,M_2^G,M_3^G)
 =m_{\mathbf1}(-1.875,\,3.125,\,1.725)
 \ \Longrightarrow\
 \frac{M_1^G}{M_2^G}=-0.6,\qquad
 \frac{M_3^G}{M_2^G}=0.552 .
\end{equation}
At one-loop common-scale running this has the same electroweak ratio as the
$\mathbf1+\mathbf{75}$ point.  Choosing the overall normalization so that
$M_2(Q_{\rm ref})\simeq120.035\TeV$ gives
$M_1(Q_{\rm ref})\simeq-38.649\TeV$ and the exact tree-level neutral gap
$\delta m_0=350\keV$ for $\mu=1.091\TeV$ and $\tan\beta=10$.
The gluino boundary condition differs, however, and enters the threshold-aware
evolution through $M_3$ and the scalar spectrum.  The value $m_{\mathbf{200}}/m_{\mathbf1}=0.05$
is an illustrative choice, not the result of a hidden-sector potential.

For prescribed unified masses $U_a\equiv M_a^G$, the boundary with
$m_{\mathbf{24}}=0$ is invertible,
\begin{equation}
 \begin{gathered}
 m_{\mathbf1}=\frac{-U_1-15U_2+40U_3}{24},\\
 m_{\mathbf{75}}=\frac{-U_1+9U_2-8U_3}{24},\qquad
 m_{\mathbf{200}}=\frac{U_1+3U_2-4U_3}{12} .
 \end{gathered}
 \label{eq:three-direction-inverse}
\end{equation}
Unlike the two-direction boundary, it permits a chosen electroweak ratio and
normalization to be maintained while $U_3$ is varied independently.  It is
therefore useful for matching a specified colored sector, but does not protect
the neutral gap against radiative corrections.

The relative coefficients and phases are hidden-sector data.  The effective
spurion parametrization does not by itself provide a dynamical vacuum-alignment
mechanism, and the three choices become distinguishable only after the colored
sector is evolved consistently.  For complex coefficients the cancellation
condition is a complex alignment constraining both magnitudes and phases; the
real plane in eq.~\eqref{eq:cancellation-plane} applies to the CP-conserving
convention used here.

If a $\mathbf{24}$ spurion also has a
scalar expectation value participating in $\mathrm{SU}(5)$ breaking, its gauge-kinetic
operator induces non-universal gauge-coupling matching. The modification of universal GUT relationship for gauge couplings is suppressed by $M_{GUT}/M_{CUT}$, which is always tiny and negligible if $M_{CUT}\sim M_{Pl}$.

\section{Phenomenological translation of the splitting}
\label{sec:phenomenology}

The reference splittings in Table~\ref{tab:reference-splittings} correspond
to different gaugino scales on the fixed-ratio high branch.

\subsection{LZ and solar-capture reference scales}
\label{subsec:phenom-scales}

For $\mu=1.091\TeV$ and $\tan\beta=10$, exact tree-level diagonalization
on the high branch gives Table~\ref{tab:updated-mapping} at the fixed ratio
\begin{equation}
 r_{\rm can}=-\frac35\frac{g_1^2(Q_{\rm ref})}{g_2^2(Q_{\rm ref})}\simeq-0.32198255.
 \label{eq:fixed-ratio-numeric}
\end{equation}

\begin{table}[!ht]
 \centering
 \small
 \begin{tabular}{@{}rcc@{}}
 \toprule
 $\delta m_0$ & $M_2(Q_{\rm ref})$ & $M_1(Q_{\rm ref})$ \\
 \midrule
 $340\keV$    & $121.885\TeV$ & $-39.245\TeV$ \\
 $350\keV$    & $120.035\TeV$ & $-38.649\TeV$ \\
 $360\keV$    & $118.262\TeV$ & $-38.078\TeV$ \\
 $371\keV$ & $116.444\TeV$ & $-37.493\TeV$ \\
 $385\keV$    & $114.136\TeV$ & $-36.750\TeV$ \\
 $400\keV$    & $111.847\TeV$ & $-36.013\TeV$ \\
 $506\keV$    & $98.683\TeV$  & $-31.774\TeV$ \\
 $566\keV$    & $92.925\TeV$  & $-29.920\TeV$ \\
 \bottomrule
 \end{tabular}
 \caption{Exact tree-level high-branch mapping in the illustrative matching
 scheme.  The masses are signed common-scale full-theory reference parameters,
 displayed to the nearest GeV, not pole masses.  Gap labels are rounded to
 $1\keV$; the $371\keV$ row uses the published $370.71\keV$ input.
 The terrestrial and solar sources are those in
 Table~\ref{tab:reference-splittings}.}
 \label{tab:updated-mapping}
\end{table}

\begin{figure}[t]
 \centering
 \includegraphics[width=0.94\textwidth]{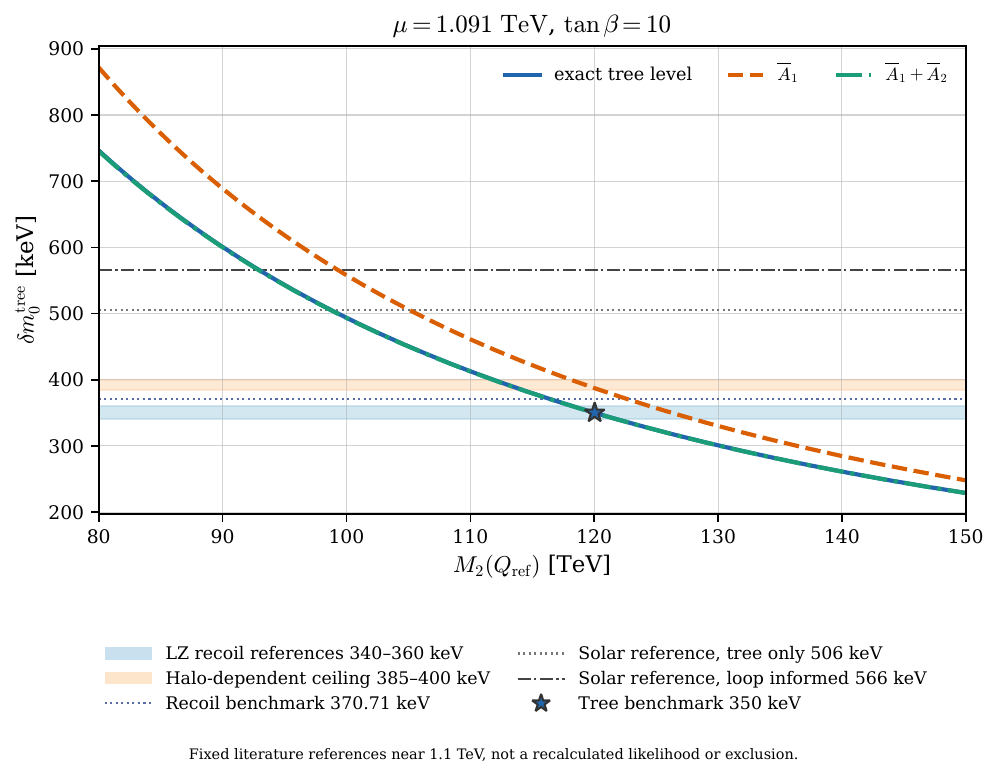}
\caption{Translation of phenomenological neutral-Higgsino splitting scales
 into the fixed-ratio electroweak-gaugino mass for
 $M_1/M_2=r_{\rm can}$.  The solid curve is the exact tree-level result and
 the dashed curves are the $\overline A_1$ and $\overline A_1+\overline A_2$ approximations.
 The star marks the $350\keV$ tree-level benchmark.  The blue band
 shows the $340$--$360\keV$ recoil reference range
 ~\cite{Freese:2026higgsino,Wu:2026higgsino}, and the separate blue line
 marks $370.71\keV$~\cite{DiMauro:2026kinematic};
 the orange band marks an approximate halo-dependent terrestrial kinematic
 ceiling~\cite{DiMauro:2026kinematic}.  The solar lines are literature-derived
 reference scales~\cite{Pospelov:2026solar}, not a combined allowed region.}
 \label{fig:scale-mapping}
\end{figure}

On the monotonic high branch, the representative recoil values between
$340$ and $371\keV$ map to $M_2\simeq116$--$122\TeV$.
This range brackets the reference choices; it is not a combined allowed band.
Figure~\ref{fig:scale-mapping} compares the exact curve with the first two
heavy-gaugino terms.

The solar-capture reference values correspond to
$M_2\lesssim98.7\TeV$ for $506\keV$ and
$M_2\lesssim92.9\TeV$ for $566\keV$.  These inequalities are conditional
on the fixed ratio, the chosen input couplings, and the branch selection.

\subsection{Solar capture and the IceCube constraint}
\label{subsec:z-overlap}

We define the neutral Weyl mass eigenstates by $\chi_i^0=N_{ij}\psi_j^0$,
with the positive-mass convention
\begin{equation}
 N^*\mathcal M_NN^\dagger
 =\diag(m_{\chi_1^0}^{\rm tree},\ldots,m_{\chi_4^0}^{\rm tree}),
 \qquad 0\leq m_{\chi_1^0}^{\rm tree}\leq\cdots\leq m_{\chi_4^0}^{\rm tree}.
 \label{eq:neutralino-takagi}
\end{equation}
Even for a real matrix, $N$ contains the phases needed to make the masses
nonnegative.  The normalized overlap for the inelastic neutral current is
\begin{equation}
 \mathcal C_{12}^{Z}\equiv
 \left|N_{13}N_{23}^*-N_{14}N_{24}^*\right| .
 \label{eq:z-overlap}
\end{equation}
This is unity for an ideal Higgsino pair, rather than the full coupling
including gauge factors.  At the $350\keV$ high-branch point,
direct diagonalization gives
\begin{equation}
 f_{H,1}\simeq0.99999897,\qquad f_{H,2}\simeq0.99999925,
 \qquad \mathcal C_{12}^{Z}\simeq0.9999991,
 \label{eq:z-overlap-numeric}
\end{equation}
where $f_{H,i}=|N_{i3}|^2+|N_{i4}|^2$.  The cancellation suppresses the
Majorana mass gap but leaves the off-diagonal $Z$ current essentially intact.
The inelastic channel relevant for solar capture therefore remains present.

A recent analysis argues that a full-density thermal Higgsino requires
$\delta m_0\gtrsim506\keV$ from inelastic slowing and
$\delta m_0\gtrsim566\keV$ after loop-induced elastic slowing
~\cite{Pospelov:2026solar}.  With those astrophysical and
transport assumptions, the $350\keV$ full-density
thermal interpretation remains in tension with solar capture, irrespective
of whether the small gap comes from decoupling or cancellation.
Solar capture, thermalization, annihilation transport and the conversion to
an IceCube limit are not recalculated here.

\section{Radiative sensitivity and spectrum consistency}
\label{sec:consistency}

\subsection{Effective coefficients across non-degenerate thresholds}
\label{subsec:rge-cancellation}

The canonical relation is an identity of the one-loop MSSM reference
evolution, not a renormalization invariant of the split theory.  Once the
scalars decouple, the beta functions of $M_1$ and $M_2$ contain the four
Higgs--Higgsino--gaugino couplings and terms proportional to $\mu$; the
equation for $\mu$ contains $M_1$ and $M_2$ in turn
~\cite{Giudice:2004tc}.  The RG equations for these mass parameters are given in
Appendix~\ref{app:split-rges}.  

For the hierarchy suggested by the tree-level reference point, the wino
is removed before the bino.  Their matching scales are
$Q_W\simeq|M_2(Q_W)|$ and $Q_B\simeq|M_1(Q_B)|$.  The sfermions, heavy-Higgs,
and gluino thresholds must also be determined from the chosen spectrum.
At tree level the sources of the two operators in
eq.~\eqref{eq:eft-dim5-operators} are~\cite{Nagata:2014wma}
\begin{equation}
 \begin{gathered}
 c_1^{(B)}(Q_B)=\frac{\widetilde g_{Yu}^{\,2}(Q_B)}{4M_1(Q_B)},\qquad
 c_2^{(B)}(Q_B)=\frac{\widetilde g_{Yd}^{\,2}(Q_B)}{4M_1(Q_B)},\\
 c_1^{(W)}(Q_W)=\frac{\widetilde g_{2u}^{\,2}(Q_W)}{4M_2(Q_W)},\qquad
 c_2^{(W)}(Q_W)=\frac{\widetilde g_{2d}^{\,2}(Q_W)}{4M_2(Q_W)} .
 \end{gathered}
 \label{eq:eft-source-matching}
\end{equation}
The denominators are signed Majorana parameters, in particular
$M_1(Q_B)<0$, not positive pole masses.
Let $C_N^{(a)}(Q)=c_1^{(a)}(Q)+c_2^{(a)}(Q)$ denote each source after
transport to a common scale $Q$ below both gauginos.  In the real convention,
the leading signed contribution is
\begin{equation}
 \Delta_s^{(5)}(Q)=v^2(Q)\bigl[C_N^{(B)}(Q)+C_N^{(W)}(Q)\bigr] .
 \label{eq:eft-neutral-gap}
\end{equation}
The cancellation must be imposed on this sum, not on masses evaluated at
different thresholds.

In the single-Higgs-doublet EFT below the heavy scalars and both
electroweak gauginos, with Higgsinos still active and bottom and tau
Yukawa couplings neglected, the one-loop equations are
\begin{equation}
 16\pi^2\frac{dc_i}{d\ln Q}
 =\left(6y_t^2+2\lambda_H-3g_2^2\right)c_i,\qquad i=1,2 ,
 \label{eq:eft-ci-rge}
\end{equation}
where 
\begin{eqnarray}
c_{1} &=& \frac{g_{1u}^{2}}{4M_{1}}+\frac{g_{2u}^{2}}{4M_{2}}\,,\\[4pt]
c_{2} &=& \frac{g_{1d}^{2}}{4M_{1}}+\frac{g_{2d}^{2}}{4M_{2}}\,,
\label{eq:c1c2}
\end{eqnarray} 
and $V(H)=\lambda_H(H^\dagger H)^2/2$~\cite{Nagata:2014wma}.

Note that $c_1$ and $c_2$ can be rewritten as 
\begin{eqnarray}
c_{1} &=& \left[\frac{g'^2}{4M_{1}}+\frac{g^2}{4M_{2}}\right]\sin^2\beta\,,\\
c_{2} &=& \left[\frac{g'^2}{4M_{1}}+\frac{g^2}{4M_{2}}\right]\cos^2\beta\,,
\end{eqnarray}
which vanishes separately during RGE evolving down to ${\cal O}(TeV)$ when the cancellation condition  
\begin{eqnarray}
\frac{g'^{2}}{4M_{1}}+\frac{g^2}{4M_{2}}=0~,
\end{eqnarray}
holds at the boundary scale. Consequently a vanishing sum imposed below the last gaugino threshold
remains zero under this homogeneous evolution. We can see that the cancellation condition is fairly robust, which can survive the RGE evolution. That is, leading order cancellation of the mass difference between the neutral components of Higgsino at the GUT scale can persist till the electroweak scale.

To survey the effects of RGE evolution, we would like to see the ratio $R_{c_i}$ ($i=1,2$) of the renormalized values to the tree level value (defined at the SUSY breaking scale $M_{\rm SUSY}$), where
\begin{equation}
R_{c_i} \equiv \frac{c_i(|\mu|)\, v^2(|\mu|)}{c_i v^2\big|_{\text{tree}}}\,, 
\label{eq:39}
\end{equation}
Here the running Higgs VEV $v$ according to Ref.~\cite{Pierce:1997} is evaluated as
\begin{equation}
v^2(Q) = \frac{4\left\{m_Z^2 + \mathrm{Re}\left[\Pi_{ZZ}^T(m_Z^2)\right]\right\}}{g'^2(Q)+g^2(Q)}\,,
\label{eq:40}
\end{equation}
where $\Pi_{ZZ}^T(m_Z^2)$ is the transverse part of the $Z$-boson self-energy in the $\overline{\mathrm{MS}}$ scheme with external
momentum set to be $p^2=m_Z^2$, and evaluated at the renormalization scale $Q$.

\begin{figure}[t]
 \centering
 \includegraphics[width=0.94\textwidth]{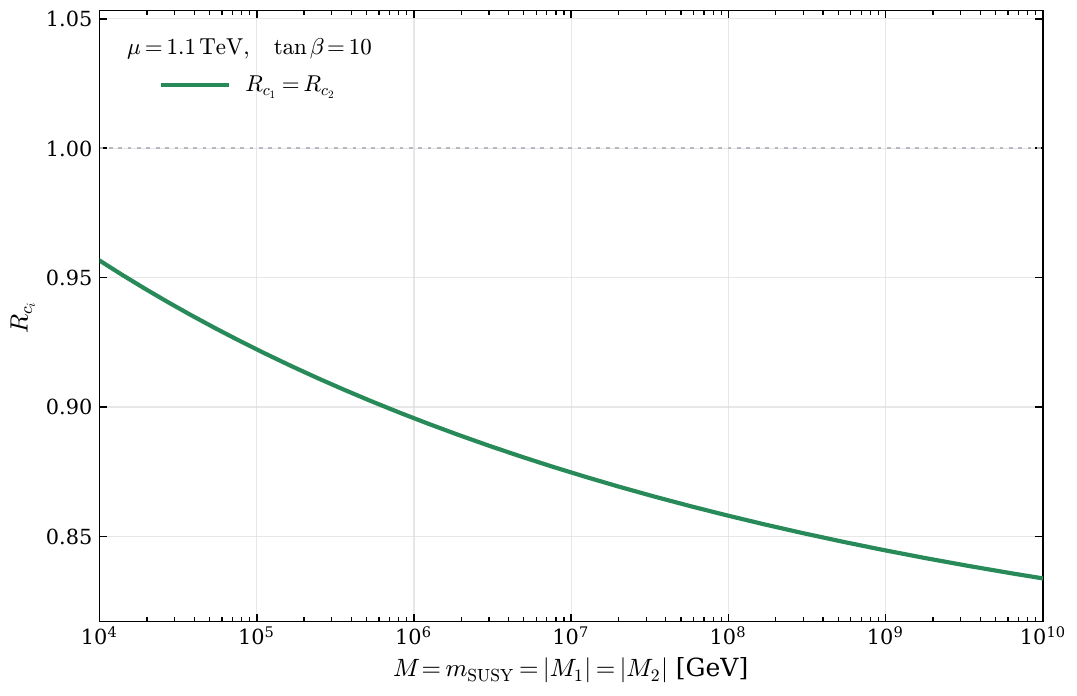}
\caption{The ratios $R$'s as functions the SUSY breaking scale. We set $\tan\beta=10$, and $\mu=1100\,\text{GeV}$. }
 \label{fig:rci-matching-scale}
\end{figure}

For later convenience, we plot the RGE evolution of $R_{c_i}$ as function of the SUSY scale in Fig.~\ref{fig:rci-matching-scale}, which we adopt the SUSY breaking scale $M_{SUSY}= M_1=M_2$. Here the two curves for $R_{c_1}=R_{c_2}$ completely coincide with tree-level matching and one-loop RGE, while threshold corrections are not included in this calculation. It can be seen that long RGE evolution decrease significantly the value of $c_i$.

Equation~\eqref{eq:eft-ci-rge} does not describe the intermediate EFT containing a dynamical bino.
The wino-generated operators in that interval require the anomalous
dimensions and matching appropriate to that field content.

In the CP-conserving Higgsino-like region, let $m_\pm^{\rm pole}$ denote
the positive pole masses continuously connected to the states with tree
eigenvalues $z_\pm$.  Define $\Delta_s^{\rm pole}=m_-^{\rm pole}-m_+^{\rm pole}$,
keeping these labels fixed across an ordering exchange.  In a fixed scheme,
the radiative corrections can be organized schematically as
\begin{equation}
 \begin{gathered}
 \Delta_s^{\rm pole}=\Delta_s^{\rm tree}
 +\delta\Delta_{\rm mass}
 +\delta\Delta_{\rm coupling}
 +\delta\Delta_{\rm evolution}\\
 \hspace{10mm}
 +\delta\Delta_{\rm finite}
 +\delta\Delta_{\rm higher}
 +\delta\Delta_{\rm self},\qquad
 \delta m_0^{\rm pole}=|\Delta_s^{\rm pole}| .
 \end{gathered}
 \label{eq:signed-correction-budget}
\end{equation}
The first two corrections that beyond the tree-level results change the threshold masses and matching vertices,
the third transports the operators, and the fourth is finite matching.
The last two denote corrections to the higher-dimensional remainder and
the low-energy self-energies.  Each term is defined relative to the same
reference parameters, with overlap subtracted.  The separation is
scheme dependent; only the consistently evaluated sum is physical.
In particular, positive gap magnitudes cannot be added term by term.

Define the reference sources and their fractional changes by
\begin{equation}
 \begin{gathered}
 C_{B,\rm ref}=\frac{g_Y^2(Q_{\rm ref})}{4M_1(Q_{\rm ref})},\qquad
 C_{W,\rm ref}=\frac{g_2^2(Q_{\rm ref})}{4M_2(Q_{\rm ref})},\\
 v^2(Q)C_N^{(a)}(Q)=v_{\rm ref}^2 C_{a,\rm ref}(1+\eta_a),
 \qquad a=B,W .
 \end{gathered}
 \label{eq:source-response-definition}
\end{equation}
These parameters collect changes of masses, couplings, matching and source
evolution; they are not predictions of a specified spectrum.  On the
canonical line $C_{B,\rm ref}=-C_{W,\rm ref}$, so
\begin{equation}
 \delta\Delta_s
 =K(\eta_W-\eta_B),\qquad
 K=\frac{m_Z^2c_W^2}{M_2(Q_{\rm ref})}\simeq52.40\MeV .
 \label{eq:source-response}
\end{equation}
The numerical value uses $M_2>0$ and the same normalization as the exact
tree matrix at the $350\keV$ high-branch point.
A differential change of about $6.7\times10^{-3}$ produces a contribution
of order $350\keV$.  A source mismatch below one percent can therefore
shift the gap by an amount comparable to the target.

\subsection{Sensitivity of the unified boundary}
\label{subsec:sensitivity}

For a positive nonzero reference gap, define
\begin{equation}
 \mathcal S_p
 =\left|\frac{p}{\delta m_0^{\rm tree}}
             \frac{\partial\delta m_0^{\rm tree}}{\partial p}\right| .
 \label{eq:sensitivity-measure}
\end{equation}
For $r=M_1/M_2<0$ this is a derivative with respect to $\ln|r|$.
When $M_2$ is varied the ratio $r$ is held fixed; when $r$ is varied $M_2$
is held fixed.  These choices distinguish a shift along the canonical
branch from a displacement across it.  The derivatives of the exact matrix
are evaluated at the unrounded $350\keV$ high-branch point. Table~\ref{tab:tree-sensitivity} gives
the resulting sensitivities, with independent finite differences used to
check the derivatives.
\begin{table}[!ht]
 \centering
 \begin{tabular}{@{}cccccc@{}}
 \toprule
 $p$ & $|r|$ & $M_2$ & $\mu$ & $\tan\beta$ & $\xi=m_{\mathbf{75}}/m_{\mathbf1}$ \\
 \midrule
 $\mathcal S_p$ & $148$ & $1.89$ & $0.894$ & $1.09$ & $158$ \\
 \bottomrule
 \end{tabular}
 \caption{Exact tree-level logarithmic sensitivities at the canonical
 $350\keV$ high-branch point.  The wino normalization is fixed when $r$ or
 $\xi$ is varied.  These are local derivatives, not pole-level tuning measures.}
 \label{tab:tree-sensitivity}
\end{table}

\begin{figure}[t]
 \centering
 \includegraphics[width=\textwidth]{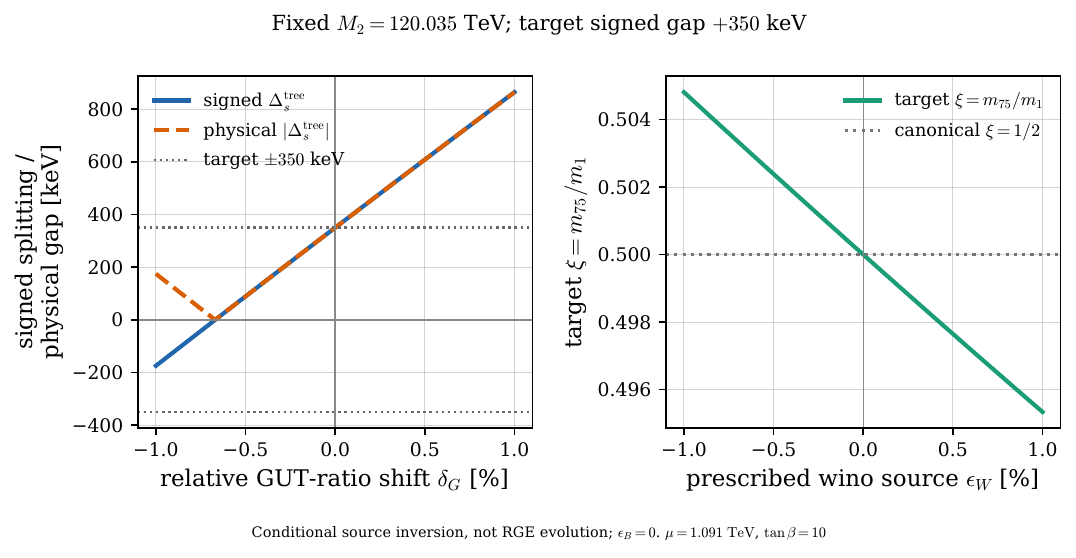}
 \caption{Sensitivity and target inversion of the reference neutralino
 matrix at fixed $M_2$, $\mu$ and $\tan\beta$.  The left panel gives the
 signed and physical tree-level gaps under a relative shift of the GUT
 ratio.  The right panel gives the coefficient $\xi$ that restores the
 signed $350\keV$ target under the prescribed wino-source deformation
 in eq.~\eqref{eq:prescribed-deformation}, with $\epsilon_B=0$.
 Representative values are listed in Table~\ref{tab:source-inversion}.  The source deformation
 is not a calculation of radiative running.}
 \label{fig:cancellation-response}
\end{figure}

The finite response is shown in Fig.~\ref{fig:cancellation-response}.
Writing $r_G=-\frac35(1+\delta_G)$, the shifts
$\delta_G=(-10^{-3},+10^{-3})$ give signed gaps $(298,402)\keV$.
For $\delta_G=(-10^{-2},+10^{-2})$ they become $(-175,864)\keV$.
The negative value denotes an exchange of the neutral-state ordering, not a
negative physical mass difference.

The hierarchy $\mathcal S_{|r|}\simeq148\gg
\mathcal S_{M_2}\simeq1.9$ shows that the small gap is governed
primarily by the relative bino--wino alignment, not the overall gaugino
normalization.  The leading estimate $\mathcal S_r\simeq K/\delta m_0$
explains this hierarchy.  On the canonical line
the first nonzero term scales as $\mu\sin2\beta/M_2^2$, whereas a ratio
displacement restores a contribution of order $K$.

These local derivatives diverge at an exact gap zero.  They measure neither
probabilities nor the dynamics selecting the spurion expectation values.
Dependence on $M_S$, $m_A$ and $M_3$, which are absent from the reference
matrix, requires the full threshold-dependent spectrum; their physical
sensitivities are not zero merely because these parameters were omitted.

\subsection{Direct inversion of the target gap}
\label{subsec:direct-target}

Write $\xi=m_{\mathbf{75}}/m_{\mathbf1}$ and hold the reference wino mass fixed by adjusting the
overall boundary normalization.  The two-direction boundary then obeys
\begin{equation}
 r_G(\xi)=\frac{1-5\xi}{1+3\xi},\qquad
 r(Q_{\rm ref})=\rho_{12}(Q_{\rm ref})r_G(\xi),\qquad
 \frac{M_3^G}{M_2^G}=\frac{1+\xi}{1+3\xi}.
 \label{eq:xi-correlated-boundary}
\end{equation}
Thus a retuning of $\xi$ also changes the correlated gluino boundary.
The normalization and ratio cannot both be fixed by one neutral-gap
measurement.  For specified Higgs and scalar parameters, that measurement
defines a curve in the two-dimensional boundary parameter space.

For a numerical example, we multiply the bino mixing entries of the
reference matrix by $\sqrt{1+\epsilon_B}$ and the wino entries by
$\sqrt{1+\epsilon_W}$, with $\epsilon_a>-1$.  Both symmetric entries
change together, while the diagonal masses remain fixed.
The resulting exact eigenvalues include the finite-gaugino remainder of
this deformed matrix.  At dimension five it gives
\begin{equation}
 C_N^{\rm def}=(1+\epsilon_B)\frac{g_Y^2(Q_{\rm ref})}{4M_1(Q_{\rm ref})}
             +(1+\epsilon_W)\frac{g_2^2(Q_{\rm ref})}{4M_2(Q_{\rm ref})}.
 \label{eq:prescribed-deformation}
\end{equation}
This is a prescribed source response, not a split-SUSY RGE solution.
It preserves the up/down coupling ratio within each source and therefore
does not replace the four independent running couplings.
For each deformation we solve the exact condition
$\Delta_s^{\rm def}=+350\keV$ at fixed $M_2$.

Table~\ref{tab:source-inversion} displays the result for
$\epsilon_B=0$ and three values of $\epsilon_W$.  The unadjusted gaps differ
substantially, whereas direct inversion restores $350\keV$ in each case.
At the canonical point,
$\partial_\xi\Delta_s^{\rm tree}\simeq110.80\MeV$ and
$\partial_{\epsilon_W}\Delta_s^{\rm def}\simeq52.50\MeV$.
The linearized response therefore gives
$\delta\xi\simeq-0.474\epsilon_W$, explaining the sign and size of the
exact solutions.  Restoring only the leading cancellation would instead
omit the finite-mass remainder that is of the same size as the target.

\begin{table}[!ht]
 \centering
 \small
 \begin{tabular}{@{}rccccc@{}}
 \toprule
 $\epsilon_W$ & $\Delta_s^{\rm def}(\xi=1/2)$ & $\xi_{\rm target}$
 & $r_G$ & $M_3^G/M_2^G$ & $\Delta_s^{\rm def}(\xi_{\rm target})$ \\
 \midrule
 $-0.01$ & $-175\keV$ & $0.504814$ & $-0.606126$ & $0.598468$ & $350\keV$ \\
 $0$     & $350\keV$  & $0.500000$ & $-0.600000$ & $0.600000$ & $350\keV$ \\
 $0.01$  & $875\keV$  & $0.495335$ & $-0.593996$ & $0.601501$ & $350\keV$ \\
 \bottomrule
 \end{tabular}
 \caption{Direct inversion for prescribed source deformations at fixed
 reference $M_2\simeq120.035\TeV$, $\mu=1.091\TeV$, $\tan\beta=10$ and
 $\epsilon_B=0$.  Calculations use the unrounded canonical high-branch mass
 and unrounded target coefficients.  Gaps are displayed to $1\keV$;
 digits in the target coefficients are not a theoretical error estimate.
 The deformations are input choices, not RGE or threshold predictions.}
 \label{tab:source-inversion}
\end{table}

Independent singlet and $\mathbf{75}$ coefficients allow a small real gap
to be retuned locally.  This parameter freedom does not provide radiative
protection.
With a $\mathbf{200}$ component, eq.~\eqref{eq:three-direction-inverse}
additionally permits the gluino boundary to be held fixed.
Applying the inversion to a physical spectrum requires evaluating
$\mathcal F_\sigma$ with the correlated gluino mass, sequential matching and
pole corrections.

\subsection{Electroweak breaking and \texorpdfstring{$B_\mu$}{Bmu}}
\label{subsec:ewsb}

The gaugino boundary condition does not determine the Higgs sector.  We use
$\mu=1.091\TeV$ and $\tan\beta=10$ as phenomenological inputs, rather than
as predictions of a particular $\mu$ or $B_\mu$ mechanism.  Our soft-potential
convention is
\begin{equation}
 V_{\rm soft}\supset m_{H_u}^2|H_u|^2+m_{H_d}^2|H_d|^2
 +\bigl(B_\mu H_u\mathbin{\cdot}H_d+\mathrm{h.c.}\bigr),
 \label{eq:higgs-soft-convention}
\end{equation}
where $H_u\mathbin{\cdot}H_d=H_u^+H_d^- -H_u^0H_d^0$, consistently with
$\epsilon_{12}=+1$.  At the matching scale, the MSSM minimization conditions,
with loop tadpoles omitted, are
\begin{align}
 \frac{m_Z^2}{2}&=
 \frac{m_{H_d}^2- m_{H_u}^2\tan^2\beta}{\tan^2\beta-1}-|\mu|^2,
 \label{eq:ewsb-one}\\
 \sin2\beta&=
 \frac{2B_\mu}{m_{H_u}^2+m_{H_d}^2+2|\mu|^2}.
 \label{eq:ewsb-two}
\end{align}

An ultraviolet completion must specify $m_{H_u}^2$, $m_{H_d}^2$, $B_\mu$,
$m_A$, the stop masses and $A_t$, and threshold corrections consistently with the Higgs
mass.  None is fixed by the cancellation plane in
eq.~\eqref{eq:cancellation-plane}.

In a threshold calculation the Higgs quartic is an additional consistency
condition.  With $V(H)=\lambda_H(H^\dagger H)^2/2$ its scalar-threshold
matching reads
\begin{equation}
 \lambda_H(M_S)=\frac{g_Y^2(M_S)+g_2^2(M_S)}{4}\cos^22\beta
              +\Delta\lambda_{H,\rm th} .
 \label{eq:higgs-quartic-matching}
\end{equation}
The correction $\Delta\lambda_{H,\rm th}$ includes finite stop, other
scalar, heavy-Higgs and scheme-conversion terms as required by the spectrum.
An infrared value extracted from the measured Higgs mass and the right-hand
side are not independent boundary inputs.  One must evolve the infrared
coupling and determine the required threshold correction, or predict the
infrared mass from specified stop and heavy-Higgs parameters
~\cite{Giudice:2004tc}.  The reference gaugino points specify neither an
electroweak-breaking solution nor a Higgs-mass benchmark.

\begin{figure}[t]
 \centering
 \includegraphics[width=0.96\textwidth]{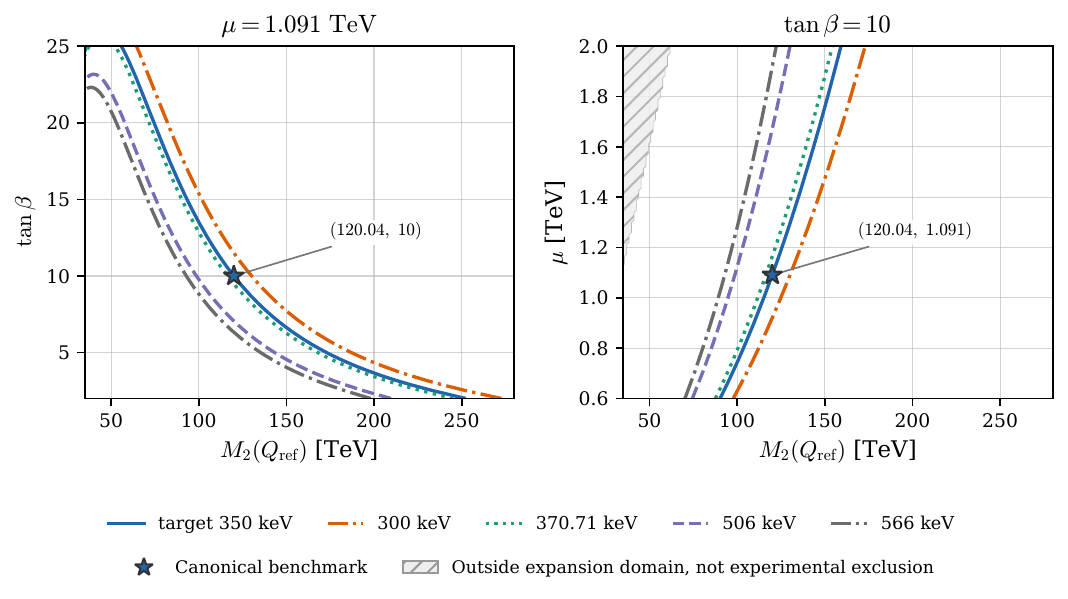}
 \caption{Fixed-ratio tree-level contours of the neutral-Higgsino gap,
 with $M_1/M_2=r_{\rm can}$.  The left panel shows the $(M_2,\tan\beta)$
 plane at $\mu=1.091\TeV$, and the right panel shows the $(M_2,\mu)$ plane
 at $\tan\beta=10$.
 The star denotes the $350\keV$ reference point.  In the right panel the
 hatched region violates $|\mu|/(|r_{\rm can}|M_2)<0.1$ and lies outside
 the controlled heavy-gaugino expansion.  It is not an experimental exclusion.
 All contours use exact diagonalization of the same common-scale tree
 matrix as the tables.}
 \label{fig:parameter-contours}
\end{figure}

The high-branch scale is also sensitive to $\tan\beta$.  For the same $\mu$
and $350\keV$ target, the fixed-ratio tree-level solutions are listed in
Table~\ref{tab:tanbeta-mapping}.  Figure~\ref{fig:parameter-contours}
extends this dependence to continuous variations of $\tan\beta$ and $\mu$.
\begin{table}[!ht]
 \centering
 \small
 \begin{tabular}{@{}rcc@{}}
 \toprule
 $\tan\beta$ & $M_2(Q_{\rm ref})$ & $M_1(Q_{\rm ref})$ \\
 \midrule
 $2$  & $252.341\TeV$ & $-81.249\TeV$ \\
 $3$  & $217.853\TeV$ & $-70.145\TeV$ \\
 $5$  & $172.919\TeV$ & $-55.677\TeV$ \\
 $10$ & $120.035\TeV$ & $-38.649\TeV$ \\
 $15$ & $93.342\TeV$  & $-30.054\TeV$ \\
 $20$ & $74.540\TeV$  & $-24.000\TeV$ \\
 \bottomrule
 \end{tabular}
 \caption{Illustrative $350\keV$ high-branch mapping as a function of
 $\tan\beta$ for $\mu=1.091\TeV$.  All mass entries are signed common-scale
 reference parameters displayed to the nearest GeV.}
\label{tab:tanbeta-mapping}
\end{table}

\subsection{Chargino spectrum and neutral-state ordering}
\label{subsec:chargino}

For $\psi^-=(\widetilde W^-,\widetilde H_d^-)^T$ and
$\psi^+=(\widetilde W^+,\widetilde H_u^+)^T$, the mass term is
$\mathcal L_{\rm mass}=-(\psi^-)^T\mathcal M_C\psi^++\mathrm{h.c.}$.
The common-scale and split-spectrum matrices are
\begin{equation}
 \begin{gathered}
 \mathcal M_C=
 \begin{pmatrix}M_2&\sqrt2\,m_{W,\rm ref}s_\beta\\
                 \sqrt2\,m_{W,\rm ref}c_\beta&\mu\end{pmatrix},\\
 \mathcal M_C^{\rm split}(Q)=
 \begin{pmatrix}M_2&\widetilde g_{2u}v/\sqrt2\\
                 \widetilde g_{2d}v/\sqrt2&\mu\end{pmatrix}_{Q}.
 \end{gathered}
 \label{eq:chargino-matrices}
\end{equation}
Here $m_{W,\rm ref}=m_Zc_W(Q_{\rm ref})\simeq79.309\GeV$ is fixed by the
neutralino reference inputs, not the measured $W$ pole mass.  The positive
tree-level chargino masses are the singular values of $\mathcal M_C$.
At the canonical ratio they give
charged-minus-lightest-neutral shifts of about $-52\MeV$ on the
$M_2\simeq11.818\TeV$ root and $-9.7\MeV$ on the asymptotic
$M_2\simeq120.035\TeV$ root.

Electroweak self-energies, which are of order
$350\MeV$ in the pure-Higgsino limit, must be added before deciding whether
the neutral state is the LSP and before computing the chargino lifetime.  The universal
gauge contribution is~\cite{Nagata:2014wma}
\begin{equation}
 \begin{gathered}
 \Delta m_+^{\rm EW}=
 \frac{\alpha_2 m_{\chi^\pm}}{4\pi}s_W^2
 f_{\rm EW}\!\left(\frac{m_Z}{m_{\chi^\pm}}\right),\\
 f_{\rm EW}(x)=2\int_0^1\!du\,(1+u)
 \ln\!\left[1+\frac{x^2(1-u)}{u^2}\right],
 \end{gathered}
 \label{eq:charged-ew-loop}
\end{equation}
with $\alpha_2=g_2^2/(4\pi)$.  The numerical estimate uses
$g_2(Q_{\rm ref})$ and $s_W(Q_{\rm ref})$, without additional evolution to
$|\mu|$.  For $m_{\chi^\pm}\simeq|\mu|=1.091\TeV$ this
gives $\Delta m_+^{\rm EW}\simeq346\MeV$, so the asymptotic branch has a
positive charged--neutral separation at this universal one-loop level.
The remaining finite-gaugino and scheme-dependent contributions must be
included in a complete electroweakino self-energy calculation to establish
\begin{equation}
 m_{\chi_1^\pm}^{\rm pole}>m_{\chi_1^0}^{\rm pole} .
 \label{eq:neutral-lsp-condition}
\end{equation}

\FloatBarrier
\section{Conclusions}
\label{sec:conclusions}

In the absence of a cancellation, a sub-MeV neutral-Higgsino splitting drives
the electroweak-gaugino scale towards $10^7\,\mathrm{GeV}$.  We have shown
that this inference changes qualitatively when the bino and wino masses have
opposite signs.  With canonical $\mathrm{SU}(5)$ normalization, the GUT-scale
ratio $M_1^G/M_2^G=-3/5$ removes the leading contribution, and homogeneous
one-loop running preserves the cancellation down to the Higgsino scale.  The
subleading finite-gaugino terms then yield a controlled tree-level branch near
$M_2=120\TeV$ for a $350\keV$ gap, $\mu=1.091\TeV$ and $\tan\beta=10$.
The convergence of the heavy-gaugino expansion supports this common-scale
result, although it does not by itself constitute a pole-mass prediction.

The required non-universal boundary condition can be realized by mixed
singlet and nonsinglet $\mathrm{SU}(5)$ representations.  Different mixtures
predict different correlated gluino boundaries, while a third independent
direction permits the colored component to be varied separately.  The small
neutral gap is considerably more sensitive to the bino--wino alignment than
to the overall gaugino scale.  Consequently, finite thresholds must be
accompanied by a corresponding retuning of the high-scale ratio.  A complete
spectrum calculation must combine sequential matching and running with a
consistent Higgs sector and neutralino and chargino pole masses evaluated in
the same renormalization scheme.

The GUT-scale cancellation addresses the hierarchy between the Higgsino and
electroweak gaugino masses, but it does not suppress the inelastic $Z$ current.
A full-density thermal Higgsino therefore remains subject to the published
solar-capture bounds under their astrophysical and transport assumptions.
Establishing a viable interpretation of the LZ event ultimately requires both
the high-scale cancellation and a pole-level spectrum consistent with these
indirect constraints.

\acknowledgments

This work was supported by the National Natural Science Foundation of China
(Grant Nos.~12447167, 12275067, 12075213 and 12335005).  Additional support
was provided by the Natural Science Foundation of Henan Province (Grant
Nos.~262300421233).  This work was also supported by the Natural Science
Foundation for Distinguished Young Scholars of Henan Province (Grant
No.~242300421046).  OpenAI Codex was used to assist with code development, figure preparation,
and language editing in this work.

\appendix
\section{Derivation of the heavy-gaugino matching}
\label{app:derivation}

Writing eq.~\eqref{eq:neutralino-matrix} in block form as
$\mathcal M_N=\left(\begin{smallmatrix}\mathcal M_g&X\\X^T&\mathcal M_H\end{smallmatrix}\right)$,
zero-momentum matching gives $\mathcal M_H-X^T\mathcal M_g^{-1}X$ at leading
order in inverse gaugino masses.
The mixing block is
\begin{equation}
 X=m_Z\begin{pmatrix}
 -s_Wc_\beta&s_Ws_\beta\\
 c_Wc_\beta&-c_Ws_\beta
 \end{pmatrix}.
 \label{eq:mixing-block}
\end{equation}
Since $\mathcal M_g=\diag(M_1,M_2)$, direct multiplication yields
\begin{equation}
 X^T\mathcal M_g^{-1}X=
 m_Z^2\left(\frac{s_W^2}{M_1}+\frac{c_W^2}{M_2}\right)
 \begin{pmatrix}c_\beta^2&-s_\beta c_\beta\\-s_\beta c_\beta&s_\beta^2\end{pmatrix},
 \label{eq:schur-product}
\end{equation}
which proves eq.~\eqref{eq:higgsino-schur}.
The perturbation is $\delta\mathcal M_H=-X^T\mathcal M_g^{-1}X$.
Its expectation value in
$h_+=(\widetilde H_d^0-\widetilde H_u^0)/\sqrt2$ is
$-\kappa(1+\sin2\beta)/2$, and that in
$h_-=(\widetilde H_d^0+\widetilde H_u^0)/\sqrt2$ is
$-\kappa(1-\sin2\beta)/2$.
Taking the absolute value of the negative unperturbed eigenvalue gives
eqs.~\eqref{eq:chi-one-mass}--\eqref{eq:neutral-splitting-leading}.

\section{Exact tree-level characteristic polynomial}
\label{app:quartic}

Taking $\det[z\mathbf1_4-\mathcal M_N]$ gives the polynomial used
for the numerical root finding in Section~\ref{sec:cancellation},
\begin{equation}
 \begin{gathered}
 P(z)=(z-M_1)(z-M_2)(z^2-\mu^2)\\
 {}-m_Z^2\bigl(z+\mu\sin2\beta\bigr)
 \left[s_W^2(z-M_2)+c_W^2(z-M_1)\right]=0 .
 \end{gathered}
 \label{eq:exact-neutralino-quartic}
\end{equation}
Its roots are signed tree-level eigenvalues in the common-scale convention
of Section~\ref{subsec:conventions}.  The corresponding physical tree-level
masses are their absolute values; neither quantity is a pole mass.

\section{Continuous retuning away from the fixed ratio}
\label{app:retuning}

At fixed $M_2$, write the displacement from the cancellation ratio as
$r=M_1/M_2=r_{\rm can}+\varepsilon_r$.  To leading order,
\begin{equation}
 \varepsilon_{r,\pm}\simeq\frac{s_W^2M_2}{c_W^4}
 \left[\mu\sin2\beta \overline A_1+\mu^2\overline A_2
 \mp\frac{\delta m_0}{m_Z^2}\right] .
 \label{eq:app-epsilon}
\end{equation}
For a singlet--$\mathbf{75}$ boundary,
\begin{equation}
 r(Q)=\rho_{12}(Q)\frac{1-5\xi}{1+3\xi},
 \qquad
 \xi(r,Q)=\frac{1-r/\rho_{12}(Q)}{5+3r/\rho_{12}(Q)} .
 \label{eq:app-xi-inversion}
\end{equation}
Thus the overall coefficient fixes the gaugino scale and $\xi$ fixes the
small residual ratio displacement.  These are common-scale MSSM reference
relations, not physical ratios in the low-energy EFT.  A pole-level
construction must include sequential matching and evolution.

Table~\ref{tab:app-retuning} gives the exact $350\keV$ roots at fixed $M_2$.
\begin{table}[!ht]
 \centering
 \small
 \begin{tabular}{@{}rccc@{}}
 \toprule
 $M_2$ & $\Delta_s^{\rm tree}$ & $r$ & $\xi=m_{\mathbf{75}}/m_{\mathbf1}$ \\
 \midrule
 $10\TeV$  & $+350\keV$ & $-0.327377$ & $0.507928$ \\
 $10\TeV$  & $-350\keV$ & $-0.327078$ & $0.507485$ \\
 $100\TeV$ & $+350\keV$ & $-0.321242$ & $0.498923$ \\
 $100\TeV$ & $-350\keV$ & $-0.317679$ & $0.493781$ \\
 \bottomrule
 \end{tabular}
 \caption{Illustrative continuous $\mathbf1+\mathbf{75}$ retuning for
 $\mu=1.091\TeV$ and $\tan\beta=10$.  Each row has a $350\keV$ physical
 tree-level gap; the sign distinguishes the two orderings.
 The displayed ratios are rounded from the exact numerical roots.}
 \label{tab:app-retuning}
\end{table}
\FloatBarrier

\section{Conventions and matching scheme}
\label{subsec:conventions}

  At the reference scale $Q_{\rm ref}=1\TeV$ we retain the
illustrative running inputs $g_1(Q_{\rm ref})=0.467451$ and
$g_2(Q_{\rm ref})=0.63811$~\cite{Antusch:2025running}.  These are SM
$\overline{\mathrm{MS}}$ inputs, not the output of an MSSM
$\overline{\mathrm{DR}}$ spectrum calculation.  They define the illustrative
one-loop reference evolution without implying pole-mass precision.

They imply
\begin{equation}
 \begin{gathered}
 \rho_{12}(Q_{\rm ref})\equiv\frac{g_1^2(Q_{\rm ref})}{g_2^2(Q_{\rm ref})}\simeq0.5366376,
 \\
 s_W^2(Q_{\rm ref})=\frac{3g_1^2(Q_{\rm ref})}{3g_1^2(Q_{\rm ref})+5g_2^2(Q_{\rm ref})}\simeq0.2435604 .
 \end{gathered}
 \label{eq:one-tev-couplings}
\end{equation}
Here $s_W=\sin\theta_W$ and $c_W=\cos\theta_W$, with $c_W^2=1-s_W^2$.
At the ideal unified boundary we impose $g_1(M_G)=g_2(M_G)=g_G$ and use
the one-loop MSSM invariant $M_a/g_a^2$ for formal reference evolution.
The parameters $M_{1,2}(Q_{\rm ref})$ are entries of a mass-independent full-theory
tree matrix, not active parameters of the physical EFT at $1\TeV$, where
both gauginos have already decoupled.  Their threshold counterparts instead
obey $Q_a\simeq|M_a(Q_a)|$.

A complete split-spectrum calculation requires conversion of the quoted
$\overline{\mathrm{MS}}$ inputs to a specified $\overline{\mathrm{DR}}$
full theory and matching onto the successive EFTs.  These effects cannot be
represented by a single multiplicative correction to the tree-level matching.

For all common-scale numerical results we take $m_Z=91.1876\GeV$ and
the weak angle in eq.~\eqref{eq:one-tev-couplings}.  To compare this matrix
with operator matching without changing its normalization, define
\begin{equation}
 v_{\rm ref}\equiv\frac{2m_Z}{\sqrt{g_Y^2(Q_{\rm ref})+g_2^2(Q_{\rm ref})}}
 \simeq248.575\GeV .
 \label{eq:reference-vev}
\end{equation}
This reference normalization is fixed by the tree-matrix inputs, rather
than extracted from the Fermi constant as a running Higgs expectation value.
In particular, replacing it by $246.22\GeV$ in only the operator contribution
would change the normalization relative to the exact matrix.

\section{Mass evolution and matching conventions}
\label{app:split-rges}

Below the scalar and heavy-Higgs thresholds but above all three gaugino
thresholds, the active fields include one scalar Higgs doublet, both
Higgsinos and all three gauginos.  We write this nonsupersymmetric EFT in
the $\MSbar$ scheme; matching to the $\DRbar$ MSSM requires finite
scheme-conversion terms.  These are not included in the reference matrix.
For real parameters and $t=\ln Q$, the one-loop equations in the convention
of ref.~\cite{Giudice:2004tc}, eqs.~(61)--(64), are
\begin{equation}
 \begin{gathered}
 16\pi^2\frac{dM_1}{dt}
 =(\widetilde g_{Yu}^2+\widetilde g_{Yd}^2)M_1
   +4\widetilde g_{Yu}\widetilde g_{Yd}\mu,\\
 16\pi^2\frac{dM_2}{dt}
 =(-12g_2^2+\widetilde g_{2u}^2+\widetilde g_{2d}^2)M_2
   +4\widetilde g_{2u}\widetilde g_{2d}\mu,\\
 16\pi^2\frac{dM_3}{dt}=-18g_3^2M_3 .
 \end{gathered}
 \label{eq:split-mass-rges}
\end{equation}
The Higgsino mass obeys
\begin{equation}
 \begin{gathered}
 16\pi^2\frac{d\mu}{dt}
 =\mu\left[-\frac32g_Y^2-\frac92g_2^2
     +\frac14(\widetilde g_{Yu}^2+\widetilde g_{Yd}^2)
     +\frac34(\widetilde g_{2u}^2+\widetilde g_{2d}^2)\right]\\
 \hspace{8mm}
 +\widetilde g_{Yu}\widetilde g_{Yd}M_1
 +3\widetilde g_{2u}\widetilde g_{2d}M_2 .
 \end{gathered}
 \label{eq:split-mu-rge}
\end{equation}
Here $g_Y=\sqrt{3/5}\,g_1$, and the supersymmetric boundary values of the
four tilde couplings are given below eq.~\eqref{eq:split-neutralino-matrix}.
Below scalar decoupling, their evolution is no longer locked to that of the
gauge couplings; both sets must be evolved together
~\cite{Giudice:2004tc,Nagata:2014wma}.

In the same interval,
\begin{equation}
 16\pi^2\frac{dg_a}{dt}=b_ag_a^3,\qquad
 (b_Y,b_2,b_3)=(15/2,-7/6,-5).
 \label{eq:split-gauge-rge}
\end{equation}
Equations~\eqref{eq:split-mass-rges}--\eqref{eq:split-gauge-rge}
do not imply a constant $M_a/g_a^2$.  Above the heavy scalar thresholds the
MSSM instead has $16\pi^2\,dM_a/dt=2b_ag_a^2M_a$, with GUT-normalized
$(b_1,b_2,b_3)=(33/5,1,-3)$, which gives the invariant used in
Section~\ref{subsec:su5}.

The displayed split equations apply only to their stated field content.
At a gaugino threshold the corresponding mass is replaced by matched
operators.  The operator basis and evolution in an interval with a remaining
gaugino must be obtained for that EFT, not by extending
eq.~\eqref{eq:eft-ci-rge} above its domain.
The common-scale tree results and prescribed source deformations in this
paper do not implement this sequential calculation.

The $\mathbf1+\mathbf{75}$ GUT boundary fixes
$M_3^G/M_2^G=(1+\xi)/(1+3\xi)$; this correlated initial condition must
be imposed in such an evolution.  Its gluino threshold is therefore not
an independent choice equal to the wino threshold.
Likewise the unified gauge matching must be satisfied at $M_G$, with any GUT
thresholds included.  A complete pole-spectrum calculation then combines this
evolution with the Higgs matching in eq.~\eqref{eq:higgs-quartic-matching}
and the neutralino and chargino self-energies
~\cite{Martin:1993yx,Pierce:1997,Fritzsche:2002,Chatterjee:2012}.

{\small
\bibliographystyle{unsrt}
\bibliography{mssm_higgsino_uv-update_references}
}

\end{document}